\documentclass[pra,twocolumn,superscriptaddress,nofootinbib,longbibliography]{revtex4-1}

\usepackage{latexsym,epsfig,epstopdf}
\usepackage{amssymb,amsmath,amsthm}

\usepackage{times}
\usepackage{color}
\usepackage{ulem}
\newcommand{\ket}[1]{\left | #1 \right\rangle}

\begin{document}

\title{An atomic test of higher-order interference}

\author{Kai Sheng Lee}
\affiliation{School of Physical and Mathematical Sciences, Nanyang Technological University, 637371 Singapore, Singapore}

\author{Zhao Zhuo}
\affiliation{School of Physical and Mathematical Sciences, Nanyang Technological University, 637371 Singapore, Singapore}

\author{Christophe Couteau}
\affiliation{Laboratory Light, nanotechnologies and nanomaterials, CNRS ERL 7004, University of Technology of Troyes, 12 rue Marie Curie, 10004 Troyes Cedex, France}

\author{David Wilkowski}
\affiliation{School of Physical and Mathematical Sciences, Nanyang Technological University, 637371 Singapore, Singapore}
\affiliation{MajuLab, International Joint Research Unit UMI 3654, CNRS, Universit\'e C\^ote d'Azur, Sorbonne Universit\'e, National University of Singapore, Nanyang Technological University, Singapore}
\affiliation{Centre for Disruptive Photonic Technologies, Nanyang Technological University, 637371 Singapore, Singapore}
\affiliation{Centre for Quantum Technologies, National University of Singapore, 117543 Singapore, Singapore}

\author{Tomasz Paterek}
\affiliation{School of Physical and Mathematical Sciences, Nanyang Technological University, 637371 Singapore, Singapore}
\affiliation{MajuLab, International Joint Research Unit UMI 3654, CNRS, Universit\'e C\^ote d'Azur, Sorbonne Universit\'e, National University of Singapore, Nanyang Technological University, Singapore}
\affiliation{Institute of Theoretical Physics and Astrophysics, Faculty of Mathematics, Physics and Informatics, University of Gda\'nsk, 80-308 Gda\'nsk, Poland}

\begin{abstract}
Canonical quantum formalism predicts that the interference pattern registered in multi-slit experiments should be a simple combination of patterns observed in two-slit experiments.
This has been linked to the validity of Born's rule and verified in precise experiments with photons as well as molecules via nuclear magnetic resonance. 
Due to the expected universal validity of Born rule, it is instructive to conduct similar tests with yet other physical systems.
Here we discuss analogs of triple-slit experiment using atoms allowing tripod energy level configuration, as realisable e.g. with alkaline-earth-like atoms.
We cover all the stages of the setup including various ways of implementing analogs of slit blockers.
The precision of the final setup is estimated and offers improvement over the previous experiments.
\end{abstract}

\maketitle

\section{Introduction}
The double-slit experiment with a single particle at a time is a landmark of quantum phenomena.
The pattern observed when two slits are open cannot be explained as a simple combination of patterns obtained from individual slits.
However, the interference fringes predicted for multi-slit quantum experiments can always be written as simple sums over the double-slit and single-slit results~\cite{Sorkin}.
Quantum mechanics is therefore termed to give rise to vanishing higher-order interference.

This observation has been linked to the validity of Born's rule and it was first tested in three-slit experiments with single photons~\cite{Weihs2010,Weihs2011}, and mimicked with nuclear magnetic resonance (NMR)~\cite{Laflamme2012}.
The figure of merit measured in these experiments is the magnitude of the triple-slit interference term (that is supposed to be vanishing) as compared to the magnitude of the double-slit fringes.
This ratio was measured down to the level of $10^{-2}$ in Ref.~\cite{Weihs2010} and in recent optics experiment it was improved to $10^{-3}$~\cite{Weihs2017} with single photons
and to $10^{-4}$ after corrections due to detector nonlinearity in the semi-classical regime~\cite{Weihs2017,Kauten2014}.
In this way it has surpassed the precision of the NMR experiment that achieved $10^{-3}$~\cite{Laflamme2012}.

It is important to mention that a small non-zero third-order interference term is actually expected in quantum multi-slit experiments, but is measurable only in the near field regime.
In essence, it is due to different boundary conditions in the experiments with many open slits versus superposition of waves emerging from many single slits~\cite{Yabuki86,RMH2012,Sinha2014,Sinha2015}.
These effects have been recently observed in Refs.~\cite{Boyd2016,Sinha2018}.

One can also question whether it is really Born rule that is being tested.
On one hand, we could take the framework of wave mechanics and replace in the intensity formula the square of the absolute of the wave function with its arbitrary power, see Ref.~\cite{Aaronson} in this context.
For this model indeed the discussed experiments put a bound on the power and verify that it has to be close to the square.
Other models, however, exist where one argues that the Born rule is preserved and yet higher-order interference is present~\cite{Quartic,Cubes}.
These models contain canonical quantum mechanics as a special case and relax other aspects of quantum formalism. 
Therefore, the experiments that rule out higher-order interference put constraints on these types of models.
Note that such generalisations of quantum mechanics have also been considered from theoretical perspective
where postulates were identified with which the models are at variance~\cite{LS2016,BLSS2017,LS2018,Zhao2018}.

Our aim here is to propose a precision test of higher-order interference using atoms.
There are few reasons that motivate this proposal.
First of all, Born rule is expected to be universally valid and yet it has not been verified with atoms.
Second, moving away from the spatial domain removes the issue with the boundary conditions~\cite{Boyd2016,Sinha2018}.
In particular, in our proposal the spatial superposition is replaced by superposition of energy eigenstates.
Third, in contradistinction to the NMR setup, atoms can be prepared in essentially pure states 
whereas NMR sample is in a highly mixed state with interference effects statistically observable for one in a million molecules.
In other words, if the complete initial NMR state was taken into account there would be almost no interference to start with.
Fourth, and different from Ref.~\cite{Laflamme2012}, we aim at providing complete analog of the spatial experiment which includes models of blocking the slits within the atomic system.
Ultimately, due to parallel processing with many atoms, the setup can also improve the results on the figure of merit from the previous experiments.
It thus has a potential of further constraining both deviations from the Born rule as well as non-quantum parameters in the generalisations of quantum mechanics.

The paper is organised as follows.
In the next section we formally introduce the concept of higher-order interference focusing on the case of three slits.
The scheme of the atomic experiment will be given in Sec.~\ref{SEC_EXP},
where in subsections we will describe state preparation, blocking of slits and final measurement.
The whole setup is essentially a Ramsey interferometer operating on atoms described by states in higher-dimensional energy subspace.
We present results of simulations of this setup in Sec.~\ref{SEC_SIM}, using fermionic strontium atoms, where the tripod scheme has been realised recently~\cite{leroux}.
The same section presents estimation of the figure of merit.
We conclude in Sec.~\ref{SEC_CONC}.

\section{Higher-order interference}
\label{SEC_INT}

Consider first the usual double-slit experiment with single photons.
Let us denote by $P_{12}$ the probability of observing the photon around a particular point on the screen if both slits are open.
Similarly one defines $P_1$ ($P_2$) as the probability, for the same point, when only the first (second) slit is open. 
The well-known experimental fact is that $P_{12} \ne P_1 + P_2$, and their difference could be thought of as quantifying quantum interference.
However, in the case of triple-slit experiment quantum mechanics predicts that the probability $P_{123}$, of detecting a photon when all three slits are open,
is a simple combination of probabilities in double-slit experiments~\cite{Sorkin}:
\begin{equation}
P_{123} = P_{12} + P_{13} + P_{23} - P_1 - P_2 - P_3.
\label{EQ_3SORKIN}
\end{equation}
Indeed one directly confirms this equation by writing each probability as squared modulus of a sum of the wave functions emitted from the corresponding slit, 
e.g. $P_{123} = |\psi_1 + \psi_2 + \psi_3|^2$, etc.
Eq.~(\ref{EQ_3SORKIN}) holds independently of the input state, i.e. uneven superpositions are also allowed, and for any observation point, making this prediction robust to experimental verification.
It is then natural to quantify the amount of hypothetical triple-slit interference as the difference:
\begin{equation}
\mathcal{S}_{3} = P_{123} - P_{12} - P_{13} - P_{23} + P_1 + P_2 + P_3.
\label{EQ_S3}
\end{equation}
In a similar manner one can define $\mathcal{S}_n$ for arbitrary number of slits,
but we note that once $\mathcal{S}_m = 0$ for some $m$, all higher-order terms, i.e. with $n > m$ also vanish~\cite{Sorkin}.
We therefore focus here on $\mathcal{S}_3$ as the first higher-order interference term that vanishes in quantum mechanics.

At first sight Eq.~(\ref{EQ_S3}) suggests that experiments with seven different combinations of open slits need to be conducted in order to estimate $\mathcal{S}_3$.
However, in order to take into account experimental background noise one subtracts from the right-hand side $P_0$, i.e. the probability of measuring the photon even if all the slits are closed~\cite{Weihs2010}:
\begin{equation}
S_3 = P_{123} - P_{12} - P_{13} - P_{23} + P_1 + P_2 + P_3 - P_0.
\label{EQ_S3_EXP}
\end{equation}
We shall refer to $S_{3}$ as the Sorkin parameter.
For comparison with previous experiments we introduce the figure of merit
defined as the ratio of the Sorkin parameter $S_3$ to the sum of double-slit interferences between all pairs of slits:
\begin{equation}
\kappa = \frac{S_3}{|S_2(12)| + |S_2(13)| + |S_2(23)|},
\label{EQ_KAPPA}
\end{equation}
where $S_{2}(jk) = P_{jk} - P_j - P_k + P_0$.
As mentioned, in the state-of-the-art experiments $\kappa$ has been estimated to the precision level of $10^{-4}$.

\section{Implementation on atoms}
\label{SEC_EXP}

\begin{figure}[!t]
	\includegraphics[scale=0.2]{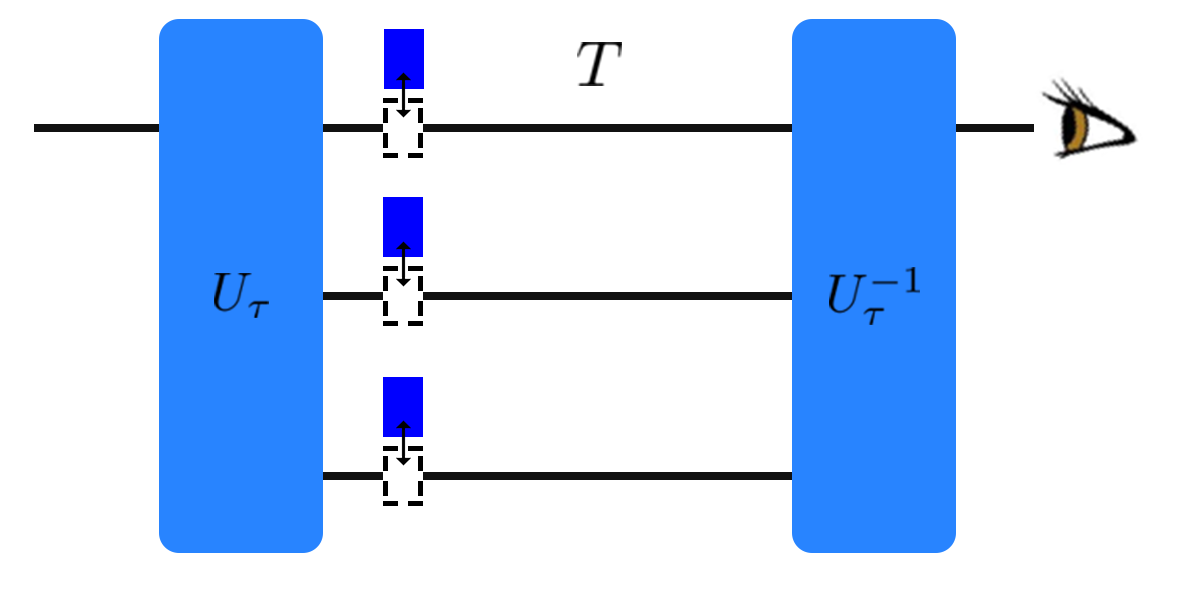}
	\centering
	\caption{Schematic of the experiment testing third-order interference.
	Temporal order is from left to right. 
	A particle is prepared in one of the three states and enters 
	the tritter box $U_\tau$ that prepares a superposition over the three possibilities.
	The boxes inside the interferometer denote the ``blockers'', i.e. they eliminate particles in states that are being blocked.
	The blockers can be put in or removed from the setup at will.
	The particle evolves inside the interferometer for time $T$.
	The second tritter closes the interferometer loop after which the probability of registering the particle in the initial state is estimated.
	In order to find $\kappa$, see Eq.~(\ref{EQ_KAPPA}), one repeats this experiment for all combinations of placing blockers in the setup.
	}
	\label{FIG_SCHEME}
\end{figure}

A generic experiment testing third-order interference is depicted and described in Fig.~\ref{FIG_SCHEME}.
In the spatial domain it is a version of Mach-Zehnder interferometer with three paths and path blockers inside the interferometer.
In the atomic setup we propose to replace the spatial paths with energy eigenstates.
The input state is then a ground state of the atom.
The role of the tritter, generalisation of the beam-splitter to three paths~\cite{Zuk-tritter}, is to prepare suitable coherent superposition of the energy states and we shall explain how this can be achieved with generalised Raman transitions in Sec.~\ref{SEC_RAMAN}.
The effect of the blockers is to proceed with the experiment only if the atoms are \emph{not} in the blocked state, without affecting coherence between the remaining states.
This is reminiscent to Leggett's ``ideal negative result'' procedures~\cite{Leggett88,Leggett08}.
We shall describe various ways of realising blockers within the atomic system in Sec.~\ref{SEC_BLOCK}.
The final stage of the setup comprises one more tritter and measurement of the probability that the atom is in the initial ground state.
Altogether the atomic setup is a Ramsey interferometer with three energy eigenstates.
Before we give the details of the implementation we show that this setup shares with the spatial setup experimentally robust vanishing third-order interference according to canonical quantum mechanics.

Consider arbitrary unitary tritter operation preparing the general superposition over the three energy eigenstates $\alpha \ket{1} + \beta \ket{2} + \gamma \ket{3}$.
Ideally, the action of a blocker is to remove from this sum the corresponding element.
For example, blocking of the first state produces atoms not in that state without changing the remaining coherences, i.e. $\beta \ket{2} + \gamma \ket{3}$. 
This output of the blocker is unnormalised to keep track of the amount of removed atoms.
Finally, the atoms are subjected to arbitrary von Neumann measurement projecting on state $\ket{m}$.
Therefore, say, $P_{123}$ is calculated as $|\langle m | (\alpha \ket{1} + \beta \ket{2} + \gamma \ket{3})|^2$ and $P_{23} = |\langle m | (\beta \ket{2} + \gamma \ket{3})|^2$, etc.
It is now straightforward to verify that $\mathcal{S}_3 = 0$ and so is the expectation for $\kappa$.
We note again that this result is independent of the initial state and the measurement.

\subsection{Initial state}
\label{SEC_INPUT}

We envisage this experiment with atoms such as strontium which in suitable energy subspace have the tripod energy diagram given in Fig.~\ref{FIG_RAMAN}.
The three lower states are stable and we shall refer to them as ground states.
We assume they have different magnetic quantum number and hence can be addressed individually with suitably polarised and arranged lasers.
We take the atoms inside external magnetic field and hence the three states have the depicted energy differences.
This energy configuration gives rise to a natural input state to the setup in Fig.~\ref{FIG_SCHEME}, i.e. ket $\ket{1}$.
We note that the initial state is practically pure after optical pumping and if needed states $\ket{2}$ and $\ket{3}$ can be obtained in essentially noiseless way via stimulated Raman adiabatic passage~\cite{STIRAP}.

\subsection{Tritter}
\label{SEC_RAMAN}

\begin{figure}[!b]
	\includegraphics[scale=0.35]{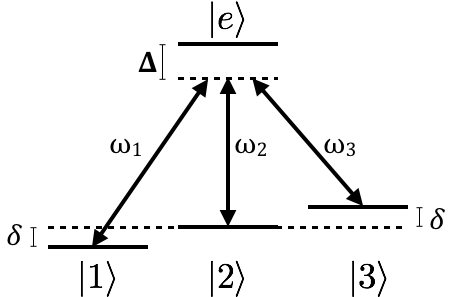}
	\centering
	\caption{The tripod energy configuration and realisation of the tritter. Starting with atoms in state $\ket{1}$ and driving with depicted lasers for time (\ref{EQ_T_TRITTER}) produces even superposition over all the ground states.
	}
	\label{FIG_RAMAN}
\end{figure}

The excited state $\ket{e}$ allows efficient preparation of coherent superpositions of the lower states via generalised Raman transitions.
Consider the atomic system and three driving lasers depicted in Fig.~\ref{FIG_RAMAN}.
The atoms are initially prepared in state $\ket{1}$. 
We take the same detuning for all three transitions, given by $\Delta$.\footnote{Throughout the paper $\hbar = 1$.}
By employing the usual assumptions, i.e. dipole coupling, rotating wave approximation, detuning much larger than the linewidth of the excited level and adiabatic elimination of the excited state,
one arrives at the effective Hamiltonian in the three-dimensional subspace of the ground states  given by (see Appendix~\ref{APP_RAMAN} for details):
\begin{equation}
\label{EQ_H_EFF}
\tilde H_{\mathrm{eff}} = - \frac{\Omega^2}{4 \Delta}
\left(
\begin{array}{ccc}
1 & 1 & 1 \\
1 & 1 & 1 \\
1 & 1 & 1
\end{array}
\right),
\end{equation}
where it is also assumed that the power and phases of the lasers are tuned such that all three Rabi frequencies are real and equal to $\Omega$.
In our notation, tilde marks operators and states in the interaction picture.
$\tilde H_{\mathrm{eff}}$ has Fourier basis as its eigenstates
\begin{eqnarray}
| \tilde \phi_k \rangle = \frac{1}{\sqrt{3}} \sum_{j = 1}^3 \eta^{k j}  | \tilde j \rangle, \quad \textrm{where} \quad \eta = e^{i \frac{2 \pi}{3}}.
\end{eqnarray}
The state $| \tilde \phi_3 \rangle$ corresponds to the eigenvalue $\tilde E_3 = - \frac{3 \Omega^2}{4 \Delta}$
whereas the other two eigenstates span degenerate subspace with eigenvalue $\tilde E_{12} = 0$.
For atoms initially prepared in state $\ket{1}$, the shortest time to evolve to an even superposition of all the three ground states is
\begin{equation}
\tau = \frac{8 \pi \Delta}{9 \Omega^2}.
\label{EQ_T_TRITTER}
\end{equation}
It is therefore this duration that we choose for implementation of the tritter.
The corresponding unitary operator of the tritter, in the Schr\"odinder picture, reads:
\begin{equation}
U_\tau = \frac{1}{\sqrt{3}}
\left(
\begin{array}{ccc}
1 & \eta e^{- i \delta \tau} & \eta e^{- i 2 \delta \tau} \\
\eta e^{i \delta \tau} & 1 & \eta e^{- i \delta \tau} \\
\eta e^{i 2 \delta \tau} & \eta e^{i \delta \tau} & 1
\end{array}
\right).
\end{equation}
Its inverse will be especially useful for the closing tritter since it leads to a maximal contrast of the interferometer. 
However, as we will discuss in more detail in Sec.~\ref{SEC_Noise}, any closing operation leads to vanishing Sorkin parameter as far as the same operation is used for all the necessary experimental realisations. 
Of course, we shall exclude trivial realisations where tritters do not create coherence in the atomic system.

\subsection{Blockers}
\label{SEC_BLOCK}

In the three-slit experiment, one can physically block the slit and in this way remove the photons that were propagating along that path. When spatial separations of the paths are absent, one has to find alternative strategies.
We wish to discuss a few ways of implementing the latter blockers in the atomic system.
Recall that ideally their role is to remove atoms in the states that are being blocked,
i.e. the output is described by an unnormalised state with unaffected coherences between unblocked states.

\subsubsection{State transfer}
\label{SEC_ELIMINATION}

Conceptually the simplest method is to transfer the blocked population into a long-lived state outside the tripod subspace.
This method could be implemented using for example metastable excited $^3 P_0$ states in alkaline-earth-like atoms. Here, the state lifetime is in the second range, i.e. well above the duration of any realistic experimental protocol. Such transitions are accessed with narrow frequency lasers.

\subsubsection{Dephasing}
\label{SEC_DEPHASING}

Since the Sorkin parameter tests certain forms of interference, it is interesting to exploit ``blocking'' realised by dephasing.
In this way the population of atoms in a blocked state does not change, but the coherences to this state are removed.
Here we show that this method indeed gives $S_3 = 0$ in an experimentally robust way, i.e. for arbitrary input state and arbitrary measurement.

Mathematically, blocking the $j$th state is realised by the dephasing map whose action on a general input density matrix $\rho$ is defined as follows:
\begin{equation}
D_j (\rho) = |j \rangle \langle j | \rho |j \rangle \langle j | + (\openone - |j \rangle \langle j |) \rho (\openone - |j \rangle \langle j |),
\end{equation}
where $\openone$ denotes identity in the ground-state subspace.
One readily verifies that this removes the off-diagonal elements in the $j$th row and $j$th column of $\rho$.
When more than one state is blocked the dephasing map $D_{jkl}(\rho)$ removes the off-diagonal elements in all the relevant rows and columns, i.e. $j$th and $k$th and $l$th in general.
Finally, we note the following identity
\begin{eqnarray}
&& \rho - D_1(\rho) - D_2(\rho) - D_3(\rho) \nonumber \\
& + & D_{12}(\rho) + D_{13}(\rho) + D_{23}(\rho) - D_{123}(\rho) = 0,
\label{EQ_D_ID}
\end{eqnarray}
which ensures that $S_3 = 0$ for arbitrary measurement.

Experimentally, one could realise controlled dephasing using lasers with different Rabi frequencies in different realisations of the experiment or by changing the duration the lasers are applied.
Consider a laser that couples only state $\ket{j}$ to the excited state.
By following the steps in Appendix~\ref{APP_RAMAN} one finds that the populations of the ground states are not modified, but state $\ket{j}$ acquires a phase $\Omega_j^2 t / 4 \Delta$.
Therefore, averaging over different realisations of the experiment averages over $\Omega_j$ or $t$ and hence removes coherences to the $j$th state.
Removal of coherences to more than one state could be done sequentially.

\subsubsection{Spontaneous emission}
\label{SEC_SPONTAN}

\begin{figure*}[!t]
	\includegraphics[scale=0.3]{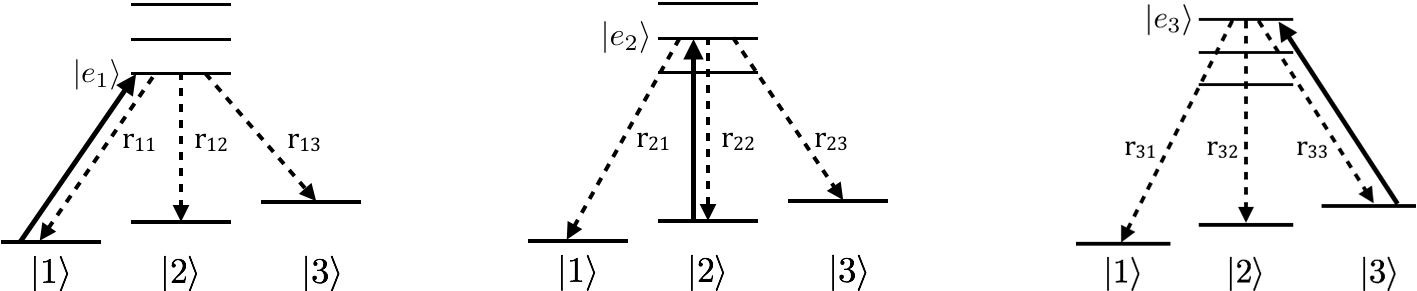}
	\centering
	\caption{``Blocking'' via spontaneous emission.
	The resonant laser moves the population of the $j$th ground state to the $j$th excited state.
	From each excited state the atoms incoherently decay to any of the ground states with probabilities $r_{kj}$ for transition from $k$th excited state to the $j$th ground state.
	}
	\label{FIG_SPONTAN}
\end{figure*}

Our last blocking procedure exploits spontaneous emission as an incoherent process removing coherences to a given state.
The process occurs naturally and hence this method is likely the simplest to implement.

We consider blocking by coupling the $j$th ground state to the $j$th excited level with the help of resonant lasers.
All the populations of the blocked states are then transferred to the corresponding excited states, from which they spontaneously decay to the ground states.
The transition probability from the $k$th excited state to the $j$th ground state is denoted by $r_{kj}$, see Fig.~\ref{FIG_SPONTAN}.
We deliberately propose to couple each ground state to a different excited state so that blocking of two and more states can be realised by simultaneously applying the corresponding lasers.
If the coupling would be through the same excited state careful analysis of coherent effects in the $\Lambda$ configuration has to be performed~\cite{EIT}.

For example, consider blocking state $\ket{2}$ as in the middle panel of Fig.~\ref{FIG_SPONTAN}.
The population of this state is brought to the excited level $\ket{e_2}$ and it then incoherently decays back.
Therefore, if $p_2 = \langle 2 | \rho | 2 \rangle$ was the probability to find the atom initially in state $\ket{2}$, after this process the atom is in this state with probability $p_2 r_{22}$.
Since spontaneous emission is incoherent, all off-diagonal elements of the density matrix in the second row and column are set to zero, i.e. $\langle j | \rho | k \rangle = 0$ for $j \ne k$ and either $j$ or $k = 2$.
The other populations increase in effect of this blocking, e.g. the portion $p_2 r_{21}$ is added incoherently to the population of the first state.
Altogether, the quantum map realising this form of blocking of the $j$th state is given by
\begin{eqnarray}
B_j(\rho) & = & (\openone - | j \rangle \langle j |) \rho (\openone - | j \rangle \langle j |) \nonumber \\
& + & \sum_{m = 1}^3 r_{jm} | m \rangle \langle j | \rho | j \rangle \langle m |.
\end{eqnarray}
Similarly, simultaneous blocking of pairs of states $j$ and $k$ is encoded by the map
\begin{eqnarray}
B_{jk}(\rho) & = & (\openone - | j \rangle \langle j | - | k \rangle \langle k |) \rho (\openone - | j \rangle \langle j | - | k \rangle \langle k |)  \\
& + & \sum_{m = 1}^3 r_{jm} | m \rangle \langle j | \rho | j \rangle  \langle m | + \sum_{m=1}^3 r_{km} | m \rangle \langle k | \rho | k \rangle \langle m | . \nonumber
\end{eqnarray}
Blocking all the states moves all the atoms to the excited states and hence the resulting density matrix after spontaneous emission is diagonal
with the $m$th diagonal entry updated as follows: $\rho_{mm} \to \sum_{j} \rho_{jj} r_{jm}$.
With this at hand one verifies that analogous identity to the one in Eq.~(\ref{EQ_D_ID}) holds after replacement $D \to B$,
and hence $S_3$ vanishes for all input states and measurements.

One can also ask whether $S_3 = 0$ if the excitation-decay cycle is repeated more times in order to more completely remove the atoms from the blocked state. 
The answer is negative. 
Already repeating this cycle three times produces non-zero Sorkin parameter. 

\subsection{Measurement}

After the blockers, the remaining atoms are allowed to freely evolve for time $T$,
after which they are measured with the combination of the closing tritter and a detector monitoring the population in state $\ket{1}$.
If the state after the free evolution is denoted by $\rho$, the probability that the detector clicks is given by
\begin{equation}
p_c = \langle 1 | U_\tau^\dagger \rho U_\tau | 1 \rangle.
\end{equation}
In other words the measurement projects onto the state $\frac{1}{\sqrt{3}}(\ket{1} + \eta e^{i \delta \tau} \ket{2} +  \eta e^{i 2 \delta \tau} \ket{3})$.

We have now covered all the elements of the atomic setup presented in Fig.~\ref{FIG_SCHEME} and we are in position to present results of its simulations.

\section{Results}
\label{SEC_SIM}

We calculate interference fringes expected in the setup for arbitrary $\delta$ and $T$, and evaluate particular noise sources and systematic effects.
The results are presented for $^{87}$Sr, the strontium fermionic isotope.

The ground state is the singlet electronic state $^1 S_0$ and the nuclear spin is $I=F = \frac{9}{2}$.
There are therefore ten ground states with different spin quantum number $m_F$.
For the present purposes we propose to use three of them with $m_F = 5/2, 7/2, 9/2$. 
We note that the experiment should be performed with a laser-cooled atomic gas loaded into a far-off-resonant optical 3D lattice operating at the magic wavelength. 
Following Ref.~\cite{leroux}, we apply a static external magnetic field of $10^{-2}\,$T giving rise to a moderate Zeeman shift of $\delta = 2\pi\times 18.2$ kHz (Land\'{e}-factor $g=-1.3\times 10^{-4}$).

To realise the tritter we propose to drive the intercombination line $^1$S$_0,F=9/2\rightarrow$ $^3$P$_1,F'=9/2$ at $689\,$nm with a linewidth of $\Gamma = 2 \pi \times 7.5\,$kHz. 
The excited state experiences a larger Zeeman shift of $\delta' = 2\pi\times 8.5\,$ MHz $=1.1\times10^3~\Gamma$ (Land\'{e}-factor $g=2/33$).
The Rabi frequency is chosen as $\Omega = 2 \pi \times 0.1$ MHz and corresponds to feasible intensities on the order $100\,$mW/cm$^2$. 
The detuning is chosen to be $\Delta = 2 \pi \times 1$ MHz, i.e. larger than the linewidth of the excited level and also smaller than the transition frequency, in agreement with the assumptions in our derivation.
It leads to tritter operation realised in a short time of $\tau = 44$ $\mu$s.

For the blocker, we focus on the method with spontaneous emission described in Sec.~\ref{SEC_SPONTAN}.
Due to the dipole selection rule $\Delta m = \pm 1, 0$, as the excited states $\ket{e_1}$, $\ket{e_2}$, $\ket{e_3}$, we chose $m'_F=7/2$ of the total nuclear spin $F' = 11/2, 9/2, 7/2$ of the electronic state $^3 P_1$, respectively.
The squared Clebsch-Gordan coefficients, connecting a ground state $\ket{j}$ to an excited state $\ket{e_k}$, give the following spontaneous emission branching ratios:
\begin{equation}
\begin{array}{ccccc}
r_{11} = \frac{1}{45}, & \quad & r_{12} = \frac{8}{45}, & \quad & r_{13} = \frac{36}{45}, \vspace{0.2cm}\\
r_{21} = \frac{32}{99}, & \quad & r_{22} = \frac{49}{99}, & \quad & r_{23} = \frac{18}{99}, \vspace{0.2cm} \\
r_{31} = \frac{36}{55}, & \quad & r_{32} = \frac{18}{55}, & \quad & r_{33} = \frac{1}{55}.
\end{array}
\end{equation}
Using similar Rabi frequency that for the tritter but at resonance, we perform $\pi$-pulse at a time $\pi/\Omega=5\,\mu$s, considerably shorter that the lifetime of the excited state $1/ \Gamma = 21\,\mu$s. 
This blocker operation should be performed after the first tritter, and followed by a free evolution longer than $1/\Gamma$ to allow complete relaxation of the excited state population.

The sequence ends with a second tritter to close the atomic interferometer, followed by a population measurement of the initial Zeeman ground substate. 
The latter could be implemented using the intercombination transition to blow away the population of the unwanted Zeeman substate followed by shot-noise-limited fluorescence measurement of the remaining population addressing the strong $^1$S$_0\rightarrow$ $^1$P$_1$ electric dipole allowed transition.

\subsection{Interferometric fringes}

\begin{figure*}
	\includegraphics[scale=0.3]{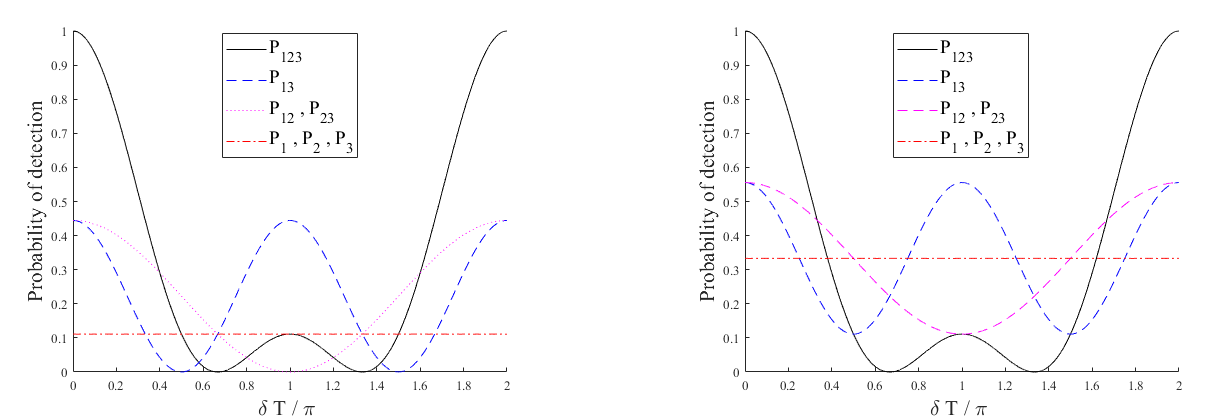}
	\centering
	\caption{Interference fringes for various combinations of blocked states.
	In the left panel the blockers are assumed to erase the corresponding part in the superposition of their input state, as described in Sec.~\ref{SEC_ELIMINATION}.
	The right panel is for the blockers realised by dephasing or spontaneous emission, as described in Sec.~\ref{SEC_DEPHASING} and Sec.~\ref{SEC_SPONTAN}.
	Both blocking methods lead to the same fringe patterns.
	}
	\label{FIG_FRINGES}
\end{figure*}

Fig.~\ref{FIG_FRINGES} shows predicted interference fringes for various combinations of blocked states.
The formulae for the probabilities depicted in the left panel are as follows:
\begin{eqnarray}
P_{123} & = & \tfrac{1}{9} \left[3 + 4 \cos(\delta \, T) + 2 \cos(2 \delta \, T) \right], \nonumber \\
P_{13} & = & \tfrac{2}{9} \left[1 + \cos(2 \delta \, T) \right], \nonumber \\
P_{12} & = & P_{23} = \tfrac{2}{9} \left[1 + \cos(\delta \, T) \right], \nonumber \\
P_1 & = & P_2 = P_3 = \tfrac{1}{9}.
\end{eqnarray}
In the right panel, the probabilities for one blocked state are shifted up by $\frac{1}{9}$ and for two blocked states by $\frac{2}{9}$.
We plot them as a function of the product of $\delta \, T$.
Therefore, for higher external magnetic field, i.e. higher $\delta$, the full fringe is observed in shorter time.
Note that the two blocking methods based on removing coherences give rise to the same interference patterns.
We also note that the interference pattern obtained by blocking the second state, $P_{13}$, is twice as fast as the patterns $P_{12}$ and $P_{23}$.
This is expected as the energy difference between states $\ket{1}$ and $\ket{3}$ is double the difference between $\ket{1}$ and $\ket{2}$ or $\ket{2}$ and $\ket{3}$.

\subsection{Noises and systematic effects}
\label{SEC_Noise}

Each box in Fig.~\ref{FIG_SCHEME} is subject to noise.
The Sorkin parameter is experimentally-friendly in a sense that it vanishes for all initial states and all measurements.
Therefore, common systematic effects, present in every experimental run, do not contribute to the estimation of $S_3$.
To illustrate this important point, let us consider the initial tritter which produces a state with different phases than required:
\begin{equation}
| \psi \rangle = \frac{1}{\sqrt{3}}(\ket{1} + \eta e^{i \delta \tau} e^{i \varphi_2} \ket{2} +  \eta e^{i 2 \delta \tau} e^{i \varphi_3} \ket{3}),
\end{equation}
where $\varphi_2$ and $\varphi_3$ are unknown.
If the tritter acts in the same way in every experimental run, this systematic effect cancels out (also if the second tritter consistently acts differently than required).
Furthermore, the phases may be random in each experimental run.
As long as their distribution is the same in all runs their average gives in effect a mixed initial state and hence the Sorkin parameter vanishes. 
Accordingly, any imperfect coherent control of the tritter, also due to spurious spontaneous emission for example, cancels out in the Sorkin parameter.

A potentially problematic situation would be a change in the remaining coherences when blocking a single state.
For example, when the first state is blocked we assumed that the remaining coherences $\rho_{23}$ and $\rho_{32}$ are not affected.
In practice, they would be affected by the AC-Stark shift, leading to additional phase $\rho_{23} \to \rho_{23} e^{i \phi}$.
One verifies that this leads to a bias of the Sorkin parameter with the magnitude given by $|S_3| = |\phi|$ in the small phase-shift approximation.
An upper bound to this phase shift is obtained by assuming that the relevant Clebsch-Gordan coefficient is equal to one.
The AC-Stark frequency shift for the $\pi$ pulse used to transfer the ground state population to the excited state is given by $\Omega^2/8\delta'=2\pi\times 150\,$Hz, leading to a relative phase shift of $5\,$mrad. 
If one is interested in improving this value so that the atomic setup improves over the previous experiments,
this phase shift can be lowered by increasing the static magnetic field (i.e. $\delta'$)
or it can be estimated using two-slit interferometer and its systematic contribution to the Sorkin parameter can be accounted in the accuracy error budget.

In our proposed setup the final measurement establishes whether the atom is in the initial state or not.
The outcome is accordingly binary and after $N$ repetitions of the experiment the estimated probability is $N p \pm \sqrt{N p(1-p)}$,
where $Np$ is the mean of the binomial distribution with $p$ being the quantum mechanical probability of atom in the initial state
and the error of $\sqrt{N p(1-p)}$ is characterised by the standard deviation of this binomial.
Since quantum mechanics predicts $S_3 = 0$, on average this value will appear after $N$ experimental runs, but with the error bar scaling as $\sqrt{N}$.
At the same time, the double-slit interferences $S_2(jk)$, see Eq.~(\ref{EQ_KAPPA}), have leading terms proportional to $N$ as quantum mechanical double-slit interference does not vanish.
Therefore, finite sample size of $N$ measurement outcomes gives rise to the value of the figure of merit $\kappa$ with error bar on the order $1 / \sqrt{N}$.

A typical cloud in cold atom experiments contains about $10^5$ independent atoms and the average over this many individual quantum systems is realised in a single experimental run.
From the parameters above we estimate that a complete single run, from producing the atomic cloud to the final measurement outcome, takes less than $1$ second.
Within a week one can therefore conduct $6 \times 10^5$ such experiments bringing the total sample size to more than $10^{10}$ and the precision of $\kappa$ down to the order of $10^{-5}$.

\section{Conclusions}
\label{SEC_CONC}

We have discussed various ways of implementing analogs of triple-slit experiments with atoms.
Such experiments could test the Born rule in previously unexplored domain and could bound non-quantum parameters in generalisations of quantum mechanics.
Due to parallel processing of many atoms in a single experimental run, 
the scheme considered offers an order of magnitude improvement in the precision of testing higher-order interference over the previous results.

\section{Acknowledgments}

This work was supported by the CQT/MoE funding grant No. R-710-002-016-271.

\appendix

\section{Tritter Hamiltonian}
\label{APP_RAMAN}

For completeness we briefly derive Eq.~(\ref{EQ_H_EFF}) of the main text.
Full Hamiltonian of the four-level system in Fig.~\ref{FIG_RAMAN}, including driving of the dipole transitions, is given by
\begin{equation}
\label{EQ_H_FULL4D}
H = 
\left(
\begin{array}{cccc}
0 & 0 & 0 & \Omega_1 \cos(\omega_1 t) \\
0 & \delta & 0 & \Omega_2 \cos(\omega_2 t)  \\
0 & 0 & 2 \delta & \Omega_3 \cos(\omega_3 t) \\
\Omega_1 \cos(\omega_1 t) & \Omega_2 \cos(\omega_2 t) & \Omega_3 \cos(\omega_3 t) & \omega_1 + \Delta
\end{array}
\right),
\end{equation}
where the phases of the lasers are tuned such that the Rabi frequencies $\Omega_j$ are real.
We move to the interaction picture with the base Hamiltonian
\begin{equation}
H_0 = \mathrm{diag}(0,\delta,2 \delta,\omega_1).
\end{equation}
One easily computes the perturbation Hamiltonian $H_1 = H - H_0$ in the interaction picture
and confirms that after the rotating wave approximation it reads
\begin{equation}
\label{EQ_H_FULL4D}
\tilde H_1 = \frac{1}{2}
\left(
\begin{array}{cccc}
0 & 0 & 0 & \Omega_1 \\
0 & 0 & 0 & \Omega_2  \\
0 & 0 & 0 & \Omega_3 \\
\Omega_1 & \Omega_2 & \Omega_3 & 2 \Delta
\end{array}
\right),
\end{equation}
where tilde marks the interaction picture.
We now impose that all the Rabi frequencies are equal to $\Omega$ and write the corresponding Schr\"odinger equation:
\begin{eqnarray}
i \dot{\tilde{c}}_j & = & \frac{\Omega}{2} \tilde c_e \quad \textrm{for} \quad j = 1,2,3, \\
i \dot{\tilde{c}}_e & = & \frac{\Omega}{2} (\tilde c_1 + \tilde c_2 + \tilde c_3) + \Delta \, \tilde c_e,
\end{eqnarray}
for the components of the four-dimensional state in the interaction picture.
By adiabatically eliminating the upper level, i.e. putting $\dot{\tilde{c}}_e = 0$, the system of equations reduces to the lower energy subspace and reads
\begin{eqnarray}
i \dot{\tilde{c}}_j & = & - \frac{\Omega^2}{4 \Delta} (\tilde c_1 + \tilde c_2 + \tilde c_3) \quad \textrm{for} \quad j = 1,2,3.
\end{eqnarray}
This corresponds to the three-dimensional effective Hamiltonian in Eq.~(\ref{EQ_H_EFF}) of the main text.

\bibliography{slits}



\end{document}